\documentclass[%
 reprint,
 amsmath,amssymb,
 aps,
]{revtex4-1}

\usepackage{graphicx}
\usepackage{dcolumn}
\usepackage{bm}

\begin{document}

\preprint{APS/123-QED}

\title{Thermal Hyper-Conductivity: radiative energy transport in hyperbolic media}

\author{J. Liu}
\author{E. Narimanov}%
 \email{enarimanov@purdue.edu.}
\affiliation{ 
School of Electrical and Computer Engineering and Birck Nanotechnology Center, Purdue University, West Lafayette, IN 47907, USA
}%

\date{\today}

\begin{abstract}
We develop a theoretical description of radiative thermal conductivity in hyperbolic metamaterials. We demonstrate a dramatic enhancement of the radiative thermal transport due to the super-singularity of the photonic density of states in hyperbolic media, leading to the radiative heat conductivity which can be comparable to the non-radiative contribution.
\begin{description}

\item[PACS numbers]
44.10.+i, 44.05.+e, 42.70.Qs

\end{description}
\end{abstract}

\maketitle


Heat conduction in bulk media is generally treated as a diffusion process in which thermal energy is carried primarily by phonons and electrons \cite {solidstate}. In these conditions, the electromagnetic thermal flux is relatively small and mostly ignored \cite {solidstate}. Even for nanoscale system where thermal conductivity is suppressed due to interface scattering, the contribution of radiative heat transport is still generally neglected \cite{JAP2003Cahill,JAP2014Cahill}. In this letter, we demonstrate that thermal conductivity of hyperbolic metamaterials (HMMs) shows a quantitatively different behavior and leads to a giant increase in radiative thermal conductance, which can be comparable with the non-radiative contribution.

The unique properties of HMMs originate from extreme dielectric tensor  anisotropy in these media. When the dielectric permittivity components in two orthogonal directions have opposite signs, the corresponding iso-frequency surface opens into a hyperboloid. For TM waves travelling in these anisotropic media, the phase space volume enclosed by the iso-frequency surface is infinite, leading to a broadband singularity of the corresponding photonic density of states \cite{SN2010}.

HMMs have recently been used to manipulate the near-field thermal radiation  \cite{PRBNefedov, APL2012Guo, OE2013Guo, PRB2013Liu, APL2013Liu, APL2013Biehs}. It has been further noted that the density of states singularity in HMMs could significantly enhance the thermal transport in the ballistic regime \cite{arXiv}. However, the ``conventional'' diffusive regime, when the HMM system size exceeds the electromagnetic absorption length, has not yet been studied. Due to the occurrence of the broadband singularity in the density of states \cite{SN2010}, the effective number of photons carrying the radiative energy flux will dramatically increase, and thus strongly enhance the radiative thermal conductivity.

 Furthermore, as metal-dielectric multilayered structures can show hyperbolic response \cite{NatMater2007Hoffman, APB2010Jacob, PNAS2012Naik, PNAS2014Naik}, the description of the radiative thermal conductivity is also relevant for practical applications. Recently, there's been a significant interest of exploring thermal transport for nanoscale multilayered devices due to promising applications in thermal barriers \cite{Science2004Costescu}, thermoelectrics \cite{HeatTrans2002Chen} and extreme ultraviolet and soft X-ray optics \cite{NanoLett2012Li,JAP2012Bozorg}. Layered materials including  W/$\rm Al_2O_3$  \cite{Science2004Costescu} and Ta/$\rm TaO_x$ laminates \cite{APL2005Ju}, Mo/Si \cite{NanoLett2012Li,JAP2012Bozorg} and Au/Si multilayers \cite{NanoLett2014Dechaumphai}, were found to have ultra-low thermal conductivities in the range of 0.33-1.5 $\rm W m^{-1} K^{-1} $. The thermal management of III-V semiconductor superlattices has also been studied \cite{JAP2006Scamarcio, IEEE1998Piprek}, as they are widely used as quantum well gain media and Bragg reflectors in quantum cascade laser devices.  Interestingly, many of these  systems are composed of alternative layers of metal and dielectric, and may display hyperbolic dispersion of the high-\textit{k} modes, with resulting broadband singularity in the bulk photonic density of states \cite{SN2010}. As a result, the anomously large photonic density of states will enhance the radiative thermal transport, and it must be taken into account in the overall thermal conductivity.

In this letter, we consider the radiative thermal transport in a uniaxial HMM along the optical axis in the $z$ direction as shown in Fig.1(a). As the radiative thermal transport through the whole structure is diffusive, we employ the Boltzman transport equation  to solve for the thermal conductivity of the radiative channel \cite{JAP2003Cahill}. 

\begin{figure} [htbp] 
\includegraphics[width=3.5in]{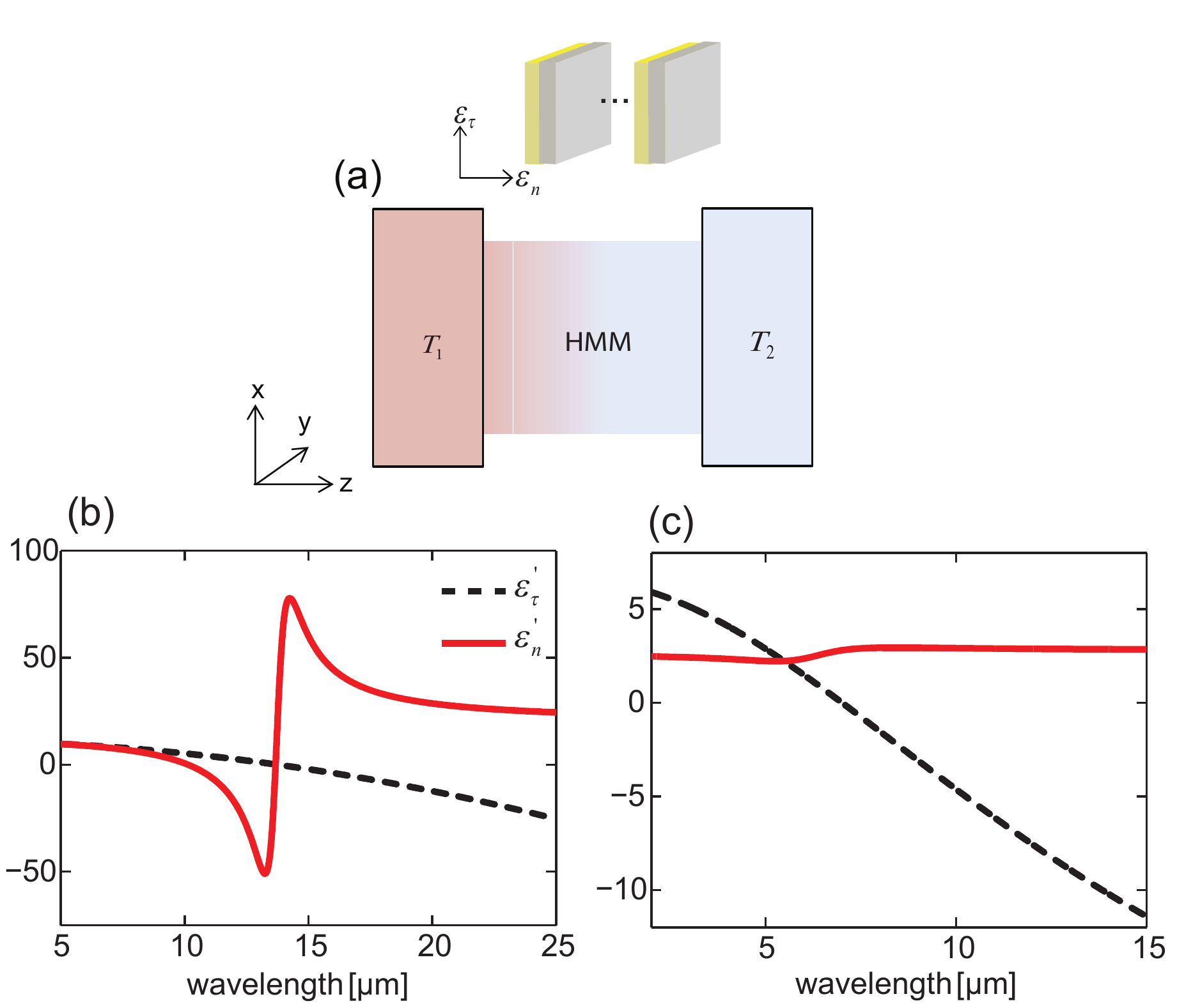}
\caption{\label{fig:epsart} Panel (a) :  Schematic of the hyperbolic metamaterials (HMM). The ends of the structure are contacting with thermal reservoirs with constant temperature $T_1$ and $T_2$. The heat flows along the optical axis in z direction. Panels (b) and (c): the real parts of permittivities of the $\rm InGaAs$ : $\rm AlInAs$ HMM (b) and the doped Si/SiO$_2$ HMM (c) calculated using the effective medium theory. In both cases, the dielectric filling ratio is $1/2$.} 
\end{figure}

Within the framework of Boltzman transport theory which is valid for the radiative heat conduction in diffusive regime, the thermal conductivity is given by \cite{JAP2003Cahill}
\begin{eqnarray}
G & =& \sum\limits_{m} \int\frac{d^3{\bf k}}{(2\pi)^3}\hbar\omega_m({\bf k}) v_{gm}^2({\bf k}) \tau_m({\bf k}) \frac{dn_B}{dT}
\label{eq:1}
\end{eqnarray}
where $v_{g}({\bf k})$ is the photon group velocity, $\tau({\bf k})$ is the thermal photon lifetime, $n_B$ is the corresponding photon occupation function, and the HMM is described by a photonic band structure $\omega({\bf k})$ with polarization  $m=[s,p]$.
As the hyperbolic mode structure and the resulting broadband density of states singularity in HMMs are only observed for 
the $p$-polarization, it is the contribution of the $p$-polarization that dominates the radiative transport in hyperbolic media,
while that of the $s$-polarization can be neglected \cite{SN2010}.  Eqn. (\ref{eq:1})  can therefore be reduced to
\begin{eqnarray}
G & = & \frac{k_B}{(2\pi)^2}\int d\omega \ I(\omega) \ F(\omega,T)
\label{eq:G}
\end{eqnarray}
where 
\begin{eqnarray}
I(\omega) =\int d k_\tau \ k_\tau \ \tau( k_\tau) \ v_g( k_\tau), 
\label{eq:I}
\end{eqnarray}
and
\begin{eqnarray}
F(\omega,T) =  (\frac {\hbar \omega}{k_B T})^2 e^{\frac {\hbar\omega}{k_BT}}/(e^{\frac {\hbar\omega}{k_BT}}-1)^2.
\label{eq:F}
\end{eqnarray}
Here $k_B$ and $\hbar$ are respectively  Boltzman and Planck constants, and $k_\tau$ is the wavevector parallel to the sample surface. Note that the original Boltzman transport theory leading to  Eqn. (\ref{eq:1}) represents essentially the leading order expression in the small parameter $1/\omega\tau \ll 1$, so the higher order terms in this parameter, arising from the integration in Eqn. (\ref{eq:G}) must be neglected.  Eqns. (\ref{eq:G})-(\ref{eq:F}) provide a closed-form analytical solution to the radiative thermal conductivity in
a hyperbolic medium.

As the first example of hyperbolic metamaterial, we choose the superlattice of highly doped semiconductor InGaAs and intrinsic AlInAs, with doped InGaAs as the medium with negative permittivity due to the electromagnetic response of free charge carriers, and insulating AlInAs as the dielectric. As  such doped semiconductor-insulator superlattices are becoming important components of modern microelectronics,  used  e.g. to reduce the leakage currents in transistor barriers,  the understanding of the different mechanisms of thermal conductivity in such materials is also important from the practical point of view. 

For the $n^+$-doped $\rm InGaAs$ : $\rm AlInAs$ HMM, the hyperbolic band covers the thermal photon energy range for the room temperature \cite{NatMater2007Hoffman},  leading to efficient thermal excitation of
high-\textit{k}  modes. We choose the thickness of each layer of the superlattice to be 50 nm which is far below the operating wavelength so that the effective medium theory (EMT) might be applicable, but also large enough so that the electronic energy quantizion can be ignored \cite{NatMater2007Hoffman}.  Under the assumption of EMT \cite{JETP1956Rytov}, the effective permittivity tensor is uniaxial and can be expressed as
\begin{eqnarray}
\epsilon_\tau & = & \frac{1}{2} \left(  \epsilon_{\rm InGaAs} + \epsilon_{\rm AlInAs} \right), \\
\epsilon_n & = & 2  \left(  \frac{1}{\epsilon_{\rm InGaAs}} + \frac{1}{\epsilon_{\rm AlInAs}} \right)^{-1} ,
\end{eqnarray}
where $\epsilon_\tau$ and $\epsilon_n$ are the effective permittivities in the directions parallel and perpendicular to plane of the layers respectively. $\epsilon_{\rm InGaAs}$ and $\epsilon_{\rm AlInAs}$ are the permittivities of the InGaAs and AlInAs layers respectively. While the permittivity of AlInAs in mid-IR is approximately constant ($\epsilon_{\rm AlInAs} = 10.23$), the permittivity of $n^+$-doped InGaAs can be expressed as \cite{NatMater2007Hoffman}  
\begin{eqnarray}
\epsilon_{\rm InGaAs} & = & \epsilon_ \infty (1-\frac{\omega_p^2}{\omega^2 + i\omega\gamma}) ,
\end{eqnarray}
 where $\epsilon_\infty = 12.15$,  $\omega_p = 1.865\times10^{14}\ \rm s^{-1}$, and $\gamma = 10^{13}\ \rm s^{-1}$.  In  Fig. 1(b) we show wavelength dependence of the resulting dielectric tensor of the $\rm InGaAs$ : $\rm AlInAs$ metamaterial.  

For our second example of HMM, we choose heavily $n^+$-doped silicon ($\sim 10^{20} \rm \ cm^{-3}$) as ``metallic'' component and SiO$_2$ as the dielectric component, with the thickness of each layer chosen to be 10 nm. As SiO$_2$ is a popular insulating material in silicon technology, the demand for understanding the precise thermal properties of these thin films is increasing. 

We will model the  permittivity of doped silicon by the Drude oscillator with parameters taken from Ref. \onlinecite{OE2008Soref}, where $\epsilon_\infty = 11.7$,  $\omega_p = 3.157\times10^{14} \rm\  s^{-1}$, and $\gamma = 1.292\times10^{14}\ \rm s^{-1}$. The permittivity of SiO$_2$ will be taken as a constant ($\epsilon_{\rm SiO_2} = 1.4$), neglecting the effect of phonon polariton resonance. Fig. 1(c) shows the effective permittivity tensor of this hyperbolic metamaterial.

The required parameters of the propagating modes in planar HMMs, such as our chosen examples,  can be obtained from the standard transfer-matrix approach \cite{Book1998Yeh},  leading to 
\begin{eqnarray}
\cos(k_nd) = \cos(k_{1z}d/2)\cos(k_{2z}d/2)\nonumber\\
-\frac{1}{2}(\frac{tk_{1z}}{k_{2z}}+\frac{k_{2z}}{tk_{1z}})\sin(k_{1z}d/2)\sin(k_{2z}d/2)
\end{eqnarray}
where the subscripts  ``1"  and ``2'' denote the parameters of the metal and the dielectric layers respectively,  $d$ is the period of the superlattice, $t=\frac {\epsilon_2}{\epsilon_1}$, and
\begin{eqnarray}
k_{(1,2)z} & =&  \sqrt{\epsilon_{(1,2)}\left(\frac{\omega}{c}\right)^2-k_\tau^2} \approx ik_\tau.
\end{eqnarray} 
The group velocity 
\begin{eqnarray}
 v_g & =& \left[ \frac{dk_n}{d\omega} \right]^{-1}
 \end{eqnarray}
 can be calculated using the Bloch wavenumber $k_n$, and the photon lifetime is determined by $\tau \approx 2/\gamma$ .  $I(\omega)$ in the Eqn. (\ref{eq:I}) is then reduced to
\begin{eqnarray}
I(\omega) & = & \frac{4\tau}{d}\frac{1}{\mid \frac{d\epsilon_r}{d\omega}\mid} \hat{F}(\epsilon_r) \label{eq:I1}
\end{eqnarray}
where
\begin{eqnarray}
\hat{F}(\epsilon_r)  =  \frac{\epsilon_r^2}{\mid\epsilon_r^2-1\mid} \int_0^{q_{\rm max}}dq\: \frac{q}{\mid 1-\cosh(q) \mid} \nonumber \\
 \times  \sqrt{1-\left[\frac{2 - \epsilon_r-\frac{1}{\epsilon_r}}{4}+\frac{2+ \epsilon_r+\frac{1}{\epsilon_r}}{4}\cosh(q)\right]^2},
\end{eqnarray}
where the dimensionless wavenumber $q\equiv k_\tau d$,  the relative permittivity $\epsilon_r\equiv \epsilon_m/\epsilon_d$, and $q_{max}$ is determined by the cut-off of $k_n$ at $\pi/ d$:
\begin{eqnarray}
q_{\rm max}  & = & {\rm arccosh} \left[ \frac{\epsilon_r+1/\epsilon_r-6}{\epsilon_r+1/\epsilon_r+2} \right].
\end{eqnarray}
 Substituting Eqn. (\ref{eq:F}) and (\ref{eq:I1}) into Eqn. (\ref{eq:G}) , we obtain
\begin{eqnarray}
G= \frac{k_B\tau}{\pi^2d}\int d\omega\  \frac {\hat{F}(\epsilon_r)}{\mid \frac{d\epsilon_r}{d\omega}\mid} \ \frac{ (\frac {\hbar \omega}{k_B T})^2 e^{\frac {\hbar\omega}{k_BT}}}{(e^{\frac {\hbar\omega}{k_BT}}-1)^2}
\end{eqnarray}
 Furthermore, $\hat{F}(\epsilon_r)$ could be  approximated as 
\begin{eqnarray}
\hat{F}(\epsilon_r) & = \frac{\pi^2}{4}
\cdot
\left\{
\begin{array}{cc}
\epsilon_r^2, &  -1 \le \epsilon_r \le 0    \\
 1,&    \epsilon_r \le -1 
 \end{array}
 \right. 
 \label{eq:Fapprox}
\end{eqnarray}
The approximation of Eqn. (\ref{eq:Fapprox}) is compared with the exact expression in Fig. 2.

\begin{figure} [htbp] 
\includegraphics[width=3in]{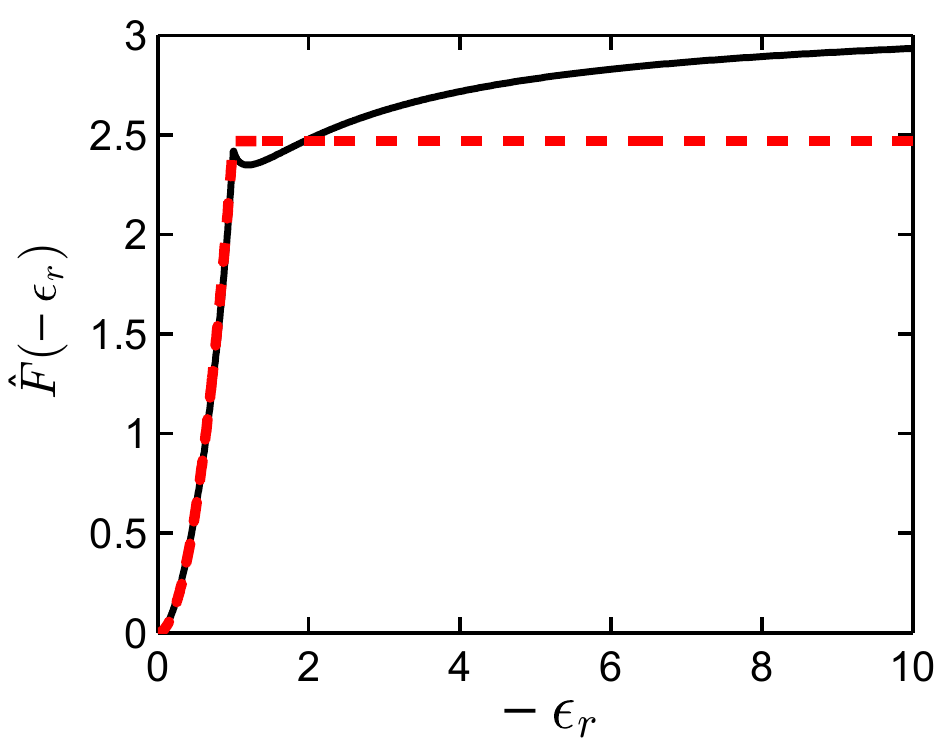}
\caption{The comparison of the exact expression of $\hat{F}(\epsilon_r)$ (black line) with the approximation of Eqn.(\ref{eq:Fapprox})(dashed red line).}
\end{figure}

The thermal conductivity G can then be represented as
\begin{eqnarray}
G = \frac{k_B\omega_p}{4d}  \frac{\omega_p}{\gamma} \frac{\epsilon_\infty}{\epsilon_d} 
\left(\frac{k_BT}{\hbar\omega_p}\right)^4 I_5\left(\frac{\hbar\omega_p}{k_BT}\right) \left[1+ \delta g\right],
\label{eq:G2}
\end{eqnarray}
where
\begin{eqnarray}
\delta g & = &  \left[ 
\left(\frac{\hbar\omega_p}{k_BT} \right)^4 \left(\frac{ \frac {\hbar \omega_1}{k_B T}}{e^{\frac {\hbar\omega_1}{k_BT}}-1}-\frac{ \frac {\hbar \omega_p}{k_B T}}{e^{\frac {\hbar\omega_p}{k_BT}}-1}+\log  \frac{1-e^{\frac {\hbar\omega_p}{k_BT}}}{1-e^{\frac {\hbar\omega_1}{k_BT}}}\right) \right. \nonumber \\
&-& 2\left(\frac{\hbar\omega_p}{k_BT} \right)^2  \left [I_3\left(\frac{\hbar\omega_p}{k_BT}\right)-I_3\left(\frac{\hbar\omega_1}{k_BT}\right) \right] \nonumber\\
& + &\left.  \left(\frac{\epsilon_d}{\epsilon_\infty}\right)^2 
I_5\left(\frac{\hbar\omega_1}{k_BT}\right)- I_5\left(\frac{\hbar\omega_1}{k_BT}\right) 
\right]  \frac {1}{I_5\left(\frac{\hbar\omega_p}{k_BT}\right)}, 
\end{eqnarray}
the integral
\begin{eqnarray}
I_n(x)& \equiv & \int_{0}^{x}\frac{t^n e^t}{(e^t-1)^2}  \ dt, 
\end{eqnarray}
and the frequency 
\begin{eqnarray}
\omega_1 \equiv  \frac{\omega_p}{\sqrt{1+\frac{\epsilon_d}{\epsilon_\infty}}}.
\end{eqnarray}

\begin{figure} [htbp] 
\includegraphics[width=3in]{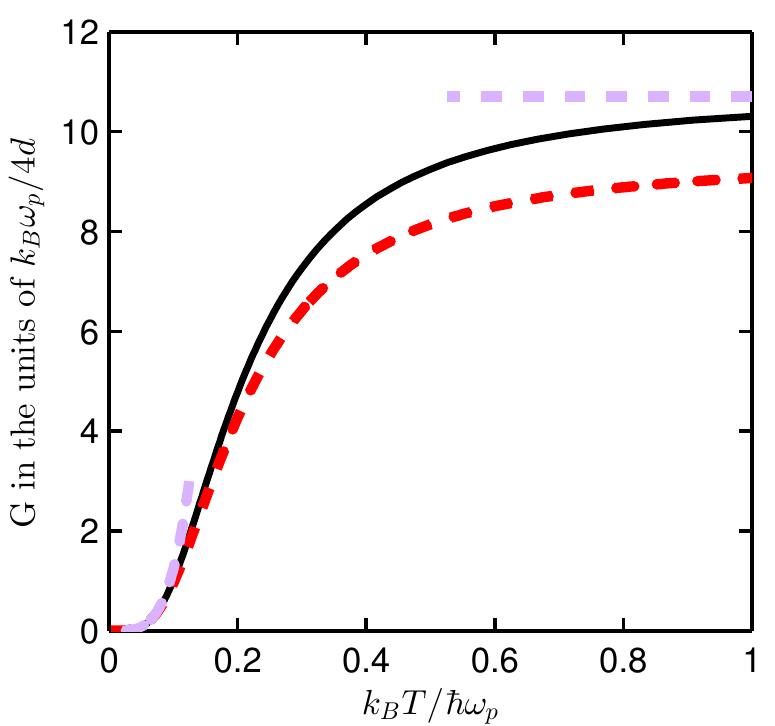}
\caption{The analytically (black line) and numerically (red dashed line) calculated thermal conductivity of the $\rm InGaAs$ : $\rm AlInAs$ HMM. $G$ is normalized by $k_B \omega_p/4d$. The purple dashed lines indicate asymptotic behavior at low and high temperatures.}
\end{figure}

 In Fig. 3  we compare the analytical result of Eqn. (\ref{eq:G2}) with the numerical calculation based on Eqn. (\ref{eq:G}).  While the deviations for the exact numerical calculations are clearly visible, our result clearly reproduces the qualitative behavior of the thermal conductivity. The  normalized themal conductivity (plotted in units of  $\frac{k_B\omega_p}{4d}$)  is increasing as $k_BT/\hbar\omega_p\ll1$ and gradually saturates as indicated by the purple dashed line when $k_BT/\hbar\omega_p\geq1$. As $k_BT\ll\hbar\omega_p$, limited number of high-k modes are excited. When temperature increases, the number of ``accessible'' high-$k$ modes increases dramatically, leading to the significant enhancement of thermal conductivity. As $k_BT$ approaches $\hbar\omega_p$, photon energy distributions fully  overlap with the super-singularity of the photonic density of states, so that nearly all the high-\textit{k} modes are now  thermally excited, and the thermal conductivity saturates.

\begin{figure} [htbp] 
\includegraphics[width=3in]{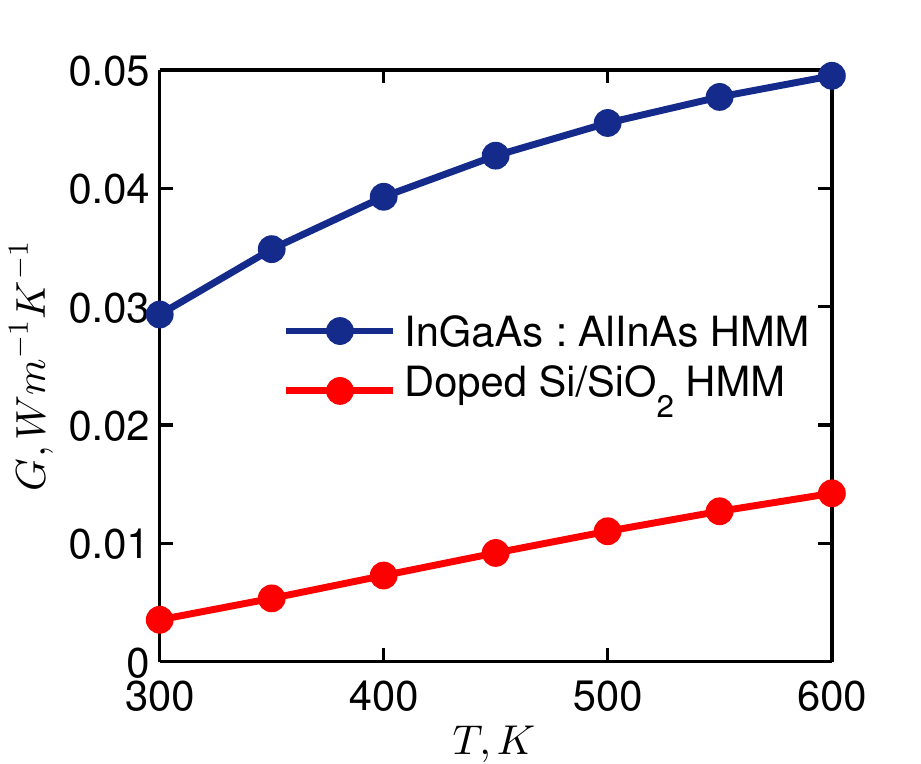}
\caption{The thermal conductivities of the $\rm InGaAs$ : $\rm AlInAs$ HMM (a) and doped Si/SiO$_2$ HMM (b) vs. temperature.}
\end{figure}

Fig.4 shows the numerically calculated thermal conductivity of our two example systems. For $\rm InGaAs$ : $\rm AlInAs$ superlattice, the onset of the saturation behavior is clearly observed, since the plasma frequency of InGaAs is significantly smaller  than that of the heavily $n^+$-doped Si. In addition, due to the longer photon lifetime in the $\rm InGaAs$ : $\rm AlInAs$  metamaterial the corresponding thermal conductivity is proportionally higher than that of the silicon-based system.

To compare our results with  the experimentally observed thermal conductivity of doped Si/SiO$_2$ HMM, we note that many research groups observed that non-radiative thermal conductivity of SiO$_2$  layer deposited on Si substrates is substantially lower than that of the bulk SiO$_2$ \cite{HeatTrans1994Goodson,JAPLEE, JAP2002Yamane,RSIChien}. In particular, the non-radiative thermal conductivity of 100 nm thick SiO$_2$ layer on Si substrate was measured to be 0.05 $\rm Wm^{-1}K^{-1}$ at 323 K \cite{HeatTrans1994Goodson,note}. For the thickness of SiO$_2$ layer down to 10 nm, the thermal conductivity will be even lower due to the pronounced effect of thermal resistance at the interface between Si and SiO$_2$. For a multilayer system based on such unit cell, the thermal conductivity of the composite would be of the same order $ \sim 0.01\rm \ Wm^{-1}K^{-1}$ \cite{NanoLett2014Dechaumphai}. From Fig. 4, we can estimate that the corresponding radiative thermal conductivity of doped Si/SiO$_2$ HMM is 0.004 $\rm Wm^{-1}K^{-1}$ at 323 K, which is of the same order as its non-radiative counterpart.

To summarize, we have presented a theoretical description of the radiative energy transport in hyperbolic metamaterials, and demonstrate that the radiative thermal conductance can be comparable with the non-radiative contribution.

\begin{acknowledgments}
This work was partially supported by NSF Center for Photonic and Multiscale Nanomaterials, ARO MURI and Gordon and Betty Moore Foundation. J. Liu was supported in part by ARO MURI Grant 56154-PH-MUR (W911NF-09-1-0539). J. Liu is grateful to the Prof. V. Shalaev for his support and helpful discussions.
\end{acknowledgments}


\begin{thebibliography}{99}

\bibitem {solidstate}
N. W. Ashcroft, and N. D. Mermin, Solid State Physics (Saunders College Publishing, Philadelphia, PA, 1976).

 \bibitem{JAP2003Cahill}%
D. G. Cahill, W. K. Ford, K. E. Goodson, G. D. Mahan, A. Majumdar, H. J. Maris, R. Merlin, and P. Sr, J. Appl. Phys. {\bf 93}, 793 (2003).

 \bibitem{JAP2014Cahill}%
 D. G. Cahill, P. V. Braun, G. Chen, D. R. Clarke, S. Fan, K. E. Goodson, P. Keblinski, W. P. King, G. D. Mahan, and A. Majumdar, Appl. Phys. Rev. {\bf 1}, 011305 (2014).
 
 \bibitem{SN2010}
 I.~I.~Smolyaninov, and E.~E.~Narimanov, Phys. Rev. Lett. {\bf 105}, 067402 (2010).
 
 \bibitem {PRBNefedov}
 I. S. Nefedov, and C. R. Simovski, Phys. Rev. B. {\bf 84}, 195459 (2011).
 
 \bibitem{APL2012Guo}%
Y. Guo, C. L. Cortes, S. Molesky, and Z. Jacob, Appl. Phys. Lett. {\bf 101}, 131106 (2012).

 \bibitem{OE2013Guo}%
Y. Guo, and Z. B. Jacob, Opt. Express {\bf 21}, 15014 (2013).

 \bibitem{PRB2013Liu}%
B. A. Liu, and S. Shen, Phys. Rev. B {\bf 87}, 115403 (2013).

 \bibitem{APL2013Liu}%
X. L. Liu, R. Z. Zhang, and Z. M. Zhang, Appl. Phys. Lett. {\bf 103}, 213102 (2013).

 \bibitem{APL2013Biehs}%
S. A. Biehs, M. Tschikin, R. Messina, and P. Ben-Abdallah, Appl. Phys. Lett. {\bf 102}, 131106 (2013).

\bibitem{arXiv}%
E. E. Narimanov, and I. I. Smolyaninov, arXiv:1109.5444 (2011).

 \bibitem{NatMater2007Hoffman}%
A. J. Hoffman, L. Alekseyev, S. S. Howard, K. J. Franz, D. Wasserman, V. A. Podolskiy, E. E. Narimanov, D. L. Sivco, and C. Gmachl, Nat. Mater. {\bf 6}, 946 (2007).

 \bibitem{APB2010Jacob}%
Z. Jacob, J. Y. Kim, G. V. Naik, A. Boltasseva, E. E. Narimanov, and V. M. Shalaev, Appl. Phys. B: Lasers Opt. {\bf 100}, 215 (2010).

 \bibitem{PNAS2012Naik}%
G. V. Naik, J. J. Liu, A. V. Kildishev, V. M. Shalaev, and A. Boltasseva, Proc. Natl. Acad. Sci.  {\bf 109}, 8834 (2012).

 \bibitem{PNAS2014Naik}%
G. V. Naik, B. Saha, J. Liu, S. M. Saber, E. A. Stach, J. M. K. Irudayaraj, T. D. Sands, V. M. Shalaev, and A. Boltasseva, Proc. Natl. Acad. Sci. {\bf 111}, 7546 (2014).

 \bibitem{Science2004Costescu}%
R. M. Costescu, D. G. Cahill, F. H. Fabreguette, Z. A. Sechrist, and S. M. George, Science {\bf 303}, 989 (2004).

 \bibitem{HeatTrans2002Chen}
G. Chen, and A. Shakouri, J. Heat Transfer {\bf 124}, 242 (2002).

 \bibitem{NanoLett2012Li}%
Z. J. Li, S. Tan, E. Bozorg-Grayeli, T. Kodama, M. Asheghi, G. Delgado, M. Panzer, A. Pokrovsky, D. Wack, and K. E. Goodson, Nano. Lett. {\bf 12}, 3121 (2012).

 \bibitem{JAP2012Bozorg}%
E. Bozorg-Grayeli, Z. J. Li, M. Asheghi, G. Delgado, A. Pokrovsky, M. Panzer, D. Wack, and K. E. Goodson, J. Appl. Phys. {\bf 112}, 083504 (2012).

 \bibitem{APL2005Ju}%
Y. S. Ju, M. T. Hung, M. J. Carey, M. C. Cyrille, and J. R. Childress, Appl. Phys. Lett. {\bf 86}, 203113 (2005).

 \bibitem{NanoLett2014Dechaumphai}%
E. Dechaumphai, D. Lu, J. J. Kan, J. Moon, E. E. Fullerton, Z. Liu, and R. Chen, Nano. Lett. {\bf 14}, 2448 (2014).

 \bibitem{JAP2006Scamarcio}%
A. Lops, V. Spagnolo, and G. Scamarcio, J. Appl. Phys. {\bf 100}, 043109 (2006).

 \bibitem{IEEE1998Piprek}%
J. Piprek, T. Troger, B. Schroter, J. Kolodzey, and C. S. Ih, IEEE Photonic Tech. Lett. {\bf 10}, 81 (1998).

 \bibitem{JETP1956Rytov}%
S. Rytov,  Sov. Phys. JETP {\bf 2}, 466 (1956).

 \bibitem{OE2008Soref}%
R. Soref, R. E. Peale, and W. Buchwald,  Opt. Express {\bf 16}, 6507 (2008).

 \bibitem{Book1998Yeh}%
P. Yeh, Optical Waves in Layered Media (Wiley-Interscience, Hoboken, New Jersey, 1998).

 \bibitem{HeatTrans1994Goodson}
K. E. Goodson, M. I. Flik, L. T. Su, and D. A. Antoniadis, J. Heat Transfer {\bf 116}, 317 (1994).

\bibitem {JAPLEE}
S. M. Lee, and D. G. Cahill, J. Appl. Phys. {\bf 81}, 2590 (1997).

 \bibitem{JAP2002Yamane}
T. Yamane, N. Nagai, S. Katayama, and M. Todoki, J. Appl. Phys. {bf 91}, 9772 (2002).

 \bibitem{RSIChien}
H. C. Chien, D. J. Yao, M. J. Huang, and T. Y. Chang,  Rev. Sci. Instrum. {\bf 79}, 054902 (2008).

 \bibitem{note}
We further note that the non-radiative thermal conductance of SiO$_2$ layers is strongly dependent on the fabrication methods. So the exact values of the observed thermal conductance of the composite are likely to vary in different experiments \cite {HeatTrans1994Goodson,JAPLEE,JAP2002Yamane,RSIChien}. 



\end{thebibliography}
\end{document}